\documentclass[letter]{aa} 

\newcommand{\kms}{km\,s$^{-1}$}

\usepackage{graphicx}
\usepackage{txfonts}
\begin{document}

   \title{High-velocity gas towards the LMC resides in the Milky Way halo
   \thanks{Based on observations obtained with the NASA/ESA
   Hubble Space Telescope, which is operated by the Space
   Telescope Science Institute (STScI) for the Association of
   Universities for Research in Astronomy, Inc., under NASA
   contract NAS5D26555.}}
   \author{
          P. Richter \inst{1,2},
          K.S. de\,Boer \inst{3},
          K. Werner \inst{4},
          \and
          T. Rauch \inst{4}
          }

   \offprints{P. Richter\\
   \email{prichter@astro.physik.uni-potsdam.de}}

   \institute{Institut f\"ur Physik und Astronomie, Universit\"at Potsdam,
   Karl-Liebknecht-Str.\,24/25, 14476 Golm, Germany
   \and
   Leibniz-Institut f\"ur Astrophysik Potsdam (AIP), An der Sternwarte 16,
   14482 Potsdam, Germany
   \and
   Argelander-Institute for Astronomy, University of Bonn,
   Auf dem H\"ugel 71, 53121 Bonn, Germany
   \and
   Institute for Astronomy and Astrophysics, Kepler Center for Astro and Particle Physics, 
   Eberhard Karls Universit\"at T\"ubingen, Sand\,1,
   72076 T\"ubingen, Germany
   }

   \date{Version 25 Sept; Received 25 September 2015; accepted xxxxxx September 2015}

 
  \abstract
  {}
  {
  To explore the origin of high-velocity gas in the direction of the
  Large Magellanic Cloud (LMC) we analyze absorption lines in the ultraviolet
  spectrum of a Galactic halo star that is located in front of the LMC
  at $d=9.2^{+4.1}_{-7.2}$ kpc distance.
  }
  {
  We study the velocity-component structure of low and intermediate
  metal ions (C\,{\sc ii}, Si\,{\sc ii}, Si\,{\sc iii}) in the spectrum 
  of RX\,J0439.8$-$6809, as obtained with the Cosmic Origins Spectrograph (COS)
  onboard the \emph{Hubble Space Telescope} (HST), and measure equivalent widths
  and column densities for these ions. We supplement our COS data with a
  \emph{Far-Ultraviolet Spectroscopic Explorer} (\emph{FUSE}) spectrum 
  of the nearby LMC star Sk\,$-$69\,59 and with H\,{\sc i} 21\,cm data 
  from the Leiden-Argentina-Bonn (LAB) survey.
  }
  {
  Metal absorption towards RX\,J0439.8$-$6809 is unambiguously detected in three 
  different velocity components near $v_{\rm LSR}=0$, $+60$, and $+150$ km\,s$^{-1}$. 
  The presence of absorption proves that all three gas components are situated in 
  front of the star, thus being located in the disk and inner halo of the Milky Way. 
  For the high-velocity cloud (HVC) at $v_{\rm LSR}=+150$ km\,s$^{-1}$ we derive an 
  oxygen abundance of [O/H$]=-0.63$ ($\sim 0.2$ solar) from the neighbouring 
  Sk\,$-$69\,59 sightline, in accordance with previous abundance measurements 
  for this HVC. From the observed kinematics we infer that the HVC hardly 
  participates in the Galactic rotation.
  }   
  {
  Our study shows that the HVC towards the LMC represents a Milky Way halo 
  cloud that traces low-column density gas with relatively low metallicity. 
  It rules out scenarios in which the HVC represents material close to the LMC 
  that stems from a LMC outflow.
  }


   \titlerunning{High-velocity gas towards the LMC resides in the MW halo}
   \authorrunning{Richter et al.}

   \maketitle


\section{Introduction}

The Milky Way is surrounded by large amounts of neutral and ionized 
gas that can be observed in absorption against bright extragalactic
background sources or in emission (e.g., in H\,{\sc i} 21cm). 
This gas is manifold in origin and chemical composition, reflecting
the various processes that determine the gas distribution around
Milky-Way type galaxies as part of cosmological structure formation
and galaxy evolution (e.g., gas infall, supernova-driven outflows, 
tidal interactions).

Extraplanar gas features in the Galactic halo manifest themselves
in absorption and emission features at high radial velocities that
are incompatible with those expected from Galactic rotation models.
Coherent gas streams with radial velocities 
$|v_{\rm LSR}|\geq 100$ km\,s$^{-1}$
are usually referred to as "high-velocity clouds" (HVCs), while those
with somewhat lower velocities ($|v_{\rm LSR}|=50-100$ km\,s$^{-1}$)
are called "intermediate-velocity clouds" 
(IVCs; see Wakker \& van Woerden 1998;
Richter 2006; Putman et al.\,2012 for recent reviews).

Previous studies have demonstrated that the majority of the large
IVCs and HVCs are located in the inner Galactic halo at distances 
$d\leq 20$ kpc (e.g., Wakker et al.\,2007, 2008; Thom et al.\,2006, 2008).
A prominent exception is the so-called Magellanic Stream, 
which represents a massive structure ($10^8-10^9$ $M_{\sun}$) 
of neutral and ionized gas at $d\sim 50$\,kpc that originates from 
the interaction of the two Magellanic Clouds orbiting the Milky Way 
(e.g., Wannier \& Wrixon 1972; 
Fox et al.\,2010, 2013, 2014; Richter et al.\,2013).

Among the various HVCs, the high-velocity cloud near $+150$ \kms\ 
in front of the LMC (hereafter referred to as HVCtwLMC)
belongs to the best studied, owing to the fact that there are many
bright LMC stars in the background 
that can be used to study the HVC with the method of
absorption spectroscopy. Previous observations indicate that HVCtwLMC is 
a relatively metal-poor ($\alpha$ abundance: $\sim 0.3$ solar; 
Lehner et al.\,2009, hereafter referred to as L09),
multi-phase gas cloud that is predominantly ionized, but that
consists of considerable small-scale structure, denser
sub-clumps, and dust (Richter et al.\,1999; Welty et al.\,2001; 
Smoker et al.\,2015). Earlier studies of HVCtwLMC have favoured a Galactic origin
(e.g., Savage \& de\,Boer 1981; de\,Boer, Morras \& Bajaja 1990;
Richter et al.\,1999), some later studies proposed that
HVCtwLMC is located close to the LMC, representing metal-enriched 
gaseous material that stems from an outflow caused by the enhanced 
star-formation activity in the Magellanic Clouds
(Stavely-Smith et al.\,2003; L09).

In this paper we demonstrate that HVCtwLMC is located in front of the star 
RX\,J0439.8$-$6809 (Fig.\,1) at a distance of $d\leq 13.3$ kpc (Werner \& Rauch 2015), 
thus proving that the gas resides within the inner halo of the Milky Way,
but is not related to a supposed LMC outflow.
The paper is organized as follows. In Sect.\,2 we briefly describe the
observations and the data analysis. In Sect.\,3 we present the results
from the spectral analysis. 
In Sect.\,4 we explore the space motion of HVCtwLMC, 
discuss its chemical composition, 
and indicate possible origins of the gas. 
We conclude or study in Sect.\,5.


 \begin{figure}[t!]
 \centering
 \includegraphics[width=7.0cm]{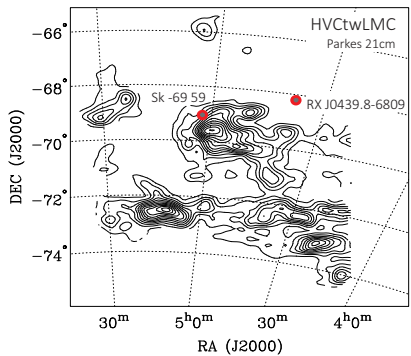}
 \caption{
 Sky position of the two background stars
 RX\,J0439.8$-$6809 (at $d <13.3$\,kpc) and Sk\,$-$69\,59 (at $d > 45$\,kpc).
 The contours display the H\,{\sc i} column density
 in the velocity range $v_{\rm LSR}=100-165$ km\,s$^{-1}$
 in steps of $2\times 10^{18}$ cm$^{-2}$
 starting at $2\times 10^{18}$ cm$^{-2}$, as observed
 in the 21cm Parkes multibeam survey of the LMC
 (adapted from Staveley-Smith et al.\,2003). The HVC presumably
 is much more extended in ionized gas, as indicated by the 
 detection of UV absorption outside the 21cm boundaries 
 (see also L09).
 }
 \label{HI21cmmap}
 \end{figure}


\section{Observations and data handling}

For our experiment we make use of archival ultraviolet (UV) spectral
data of the extremely hot white dwarf (WD) RX\,J0439.8$-$6809. 
Since its discovery in 1994 as ROSAT supersoft X-ray source, 
RX\,J0439.8$-$6809 was regarded as a WD located in the LMC 
(Greiner et al. 1994).
A recent spectral analysis of RX\,J0439.8$-$6809 
by Werner \& Rauch (2015) based on COS data reveals, however, 
that the star has a spectroscopic distance of just $d=9.2^{+4.1}_{-7.2}$,
thus being located within the Milky Way halo $\sim 5.6$ kpc below the
Galactic disk. 
The distance and the position (at $l=279.9$, $b=-37.1$) in front of 
the LMC make RX\,J0439.8$-$6809 the ideal target to constrain the distance of 
intermediate- and high-velocity gas in the direction of the Magellanic Clouds. 

We here use the same COS data set as presented in Werner \& Rauch (2015). 
RX\,J0439.8$-$6809 was observed
with \emph{HST}/COS using the G130M grating that covers the 
wavlength range between $\lambda =1150-1450$ \AA\,at a spectral resolution
of $R\approx 15\,000-20\,000$ ($15-20$ km\,s$^{-1}$) FWHM.
The total exposure time was 14\,080\,s.
We retrieved the original {\tt x1d} fits files (processed with the
CALCOSpipeline v2.17.3) from the MAST archive and coadded the individual
sciences exposures following the procedure described in Richter et al.\,(2015).
The signal-to-noise (S/N) in the final coadded spectrum is $\sim 13$ per
resolution element at $1300$\,\AA.

We supplement our COS data with a \emph{FUSE} Far-UV spectrum 
of the LMC Wolf-Rayet star Sk\,$-$69\,59 
(at $l=280.1$, $b=-34.8$, $\sim 2.5$ degrees away from RX\,J0439.8$-$6809)
to provide a reference spectrum that contains 
all relevant absorption components along the line of sight towards 
(and through) the LMC. 
Also the \emph{FUSE} data were retrieved from the MAST archive and 
were reduced in a way similar as described in Richter et al.\,(2013).

The sky positions of RX\,J0439.8$-$6809 and Sk\,$-$69\,59 are plotted 
in Fig.\,1 together with an H\,{\sc i} 21cm map of the HVCtwLMC
(from Staveley-Smith et al.\,2003).
For the spectral analysis of the COS and \emph{FUSE} data we used the 
custom-written {\tt span} software package, that is based on ESO-MIDAS 
and its {\tt fitlyman} spectral library (Fontana \& Ballester 1995).
Equivalent widths and column densities were derived from a direct pixel
integration of the absorption profiles using the apparent-optical depth 
(AOD) method (Savage \& Sembach 1991). Atomic data have been adopted from
the compilation by Morton (2003).

To derive the H\,{\sc i} column densities (or limits) in 
the direction of the two background sources we consider publicly available 
21\,cm spectral data from the LAB survey 
(Kalberla et al.\,2005) and the relation
$N($H\,{\sc i}$)=1.823\times10^{18}\,{\rm cm}^{-2}\,
\int^{v_{\rm max}}_{v_{\rm min}}\,T_{\rm B}\,{\rm d}v$.


\begin{table}[b!]
\caption[]{HVC absorption- and emission-line measurements}
\begin{small}
\begin{tabular}{lrrrr}
\hline
Ion & $\lambda_0$ [\AA] & Instrument & $W_{\lambda}$ [m\AA] & log $N$ \\
\hline
\hline
\multicolumn{5}{c}{ RX\,J0439.8$-$6809; $v_{\rm LSR}=+150$ km\,s$^{-1}$}\\
\hline
C\,{\sc ii}   & $1334.53$ & COS & $116\pm12$ & $>13.90$ \\
Si\,{\sc iii} & $1206.50$ & COS & $177\pm12$ & $>13.14$ \\
Si\,{\sc ii}  & $1260.42$ & COS & $98\pm11$  & $>12.89$ \\
Si\,{\sc ii}  & $1193.29$ & COS & $78\pm11$  & $13.10\pm0.12$ \\
Si\,{\sc ii}  & $1190.42$ & COS & $10\pm4$   & $13.04\pm0.16$ \\
H\,{\sc i}    & 21\,cm    & LAB &            & $<18.73$ \\
\hline
\multicolumn{5}{c}{Sk\,$-$69\,59; $v_{\rm LSR}=+170$ km\,s$^{-1}$}\\
\hline
O\,{\sc i}    & $1039.23$ & \emph{FUSE} & $78\pm6$  & $15.12\pm0.04$ \\
N\,{\sc i}    & $1134.98$ & \emph{FUSE} & $21\pm4$  & $13.68\pm0.07$ \\
N\,{\sc ii}   & $1083.99$ & \emph{FUSE} & $156\pm7$ & $\geq 14.25$ \\
Fe\,{\sc ii}  & $1144.94$ & \emph{FUSE} & $63\pm5$  & $13.83\pm0.05$ \\
P\,{\sc ii}   & $1152.82$ & \emph{FUSE} & $\leq 12$ & $\leq 12.10$ \\
H\,{\sc i}    & 21\,cm    &        LAB  &           & $19.06\pm0.06$ \\
\hline
\end{tabular}
\end{small}
\noindent
\label{tabCol}
\\
\end{table}


\section{Results from the spectral analysis}

In the upper panel of Fig.\,2
we show the (interstellar) absorption profiles 
of C\,{\sc ii} $\lambda 1334.53$, Si\,{\sc iii} $\lambda 1206.50$, and 
Si\,{\sc ii} $\lambda 1260.42$ in the COS spectrum of RX\,J0439.8$-$6809 
plotted on a Local Standard of Rest (LSR) velocity scale together
with the LAB 21\,cm emission profile. 
Absorption in these sight lines is detected in 
three main absorption components centered
at $v_{\rm LSR}=0,+60$ and $+150$ km\,s$^{-1}$ (dotted lines). 
The intermediate-velocity
component at $v_{\rm LSR}=+60$ km\,s$^{-1}$ blends with the very strong
local disk absorption at zero velocities, as observed along many LMC
sightlines (see Danforth et al.\,2002). 
The high-velocity gas at $v_{\rm LSR}=+150$ km\,s$^{-1}$ 
is well separated from the other velocity components. 
H\,{\sc i} 21\,cm emission (lower panel) is seen only near 
zero velocities (Galactic disk) and near $v_{\rm LSR}=+250$ km\,s$^{-1}$,
the latter component coming from neutral gas in the LMC {\it behind} 
RX\,J0439.8$-$6809, gas not present in absorption. 

In the lower panel of Fig.\,2 we show the absorption profiles 
of O\,{\sc i} $\lambda 13.3.17$, N\,{\sc i} $\lambda 1134.98$, and 
Fe\,{\sc ii} $\lambda 1144.94$ in the \emph{FUSE} spectrum of Sk\,$-$69\,59.
Here, the high-velocity component is shifted towards $v_{\rm LSR}=+170$ 
km\,s$^{-1}$, while no intermediate-velocity component is present. 
Absorption at $v_{\rm LSR}=+200$ to $+270$ km\,s$^{-1}$ stems from gas 
within the LMC. In contrast to RX\,J0439.8$-$6809, weak 21\,cm emission 
is detected in the HVC component towards Sk\,$-$69\,59. 
but the emission is slightly shifted to velocities near $v_{\rm LSR}=+150$
km\,s$^{-1}$. We attribute the velocity shift to beam-smearing effects 
in the LAB 21cm data that have a spatial resolution of $\sim 36 \arcmin$.

For the HVCtwLMC gas we have derived ion equivalent widths and 
column densities from the COS, \emph{FUSE}, and LAB data. 
The results are summarized in Table 1.


 \begin{figure}[t!]
 \centering
 \includegraphics[width=5.0cm]{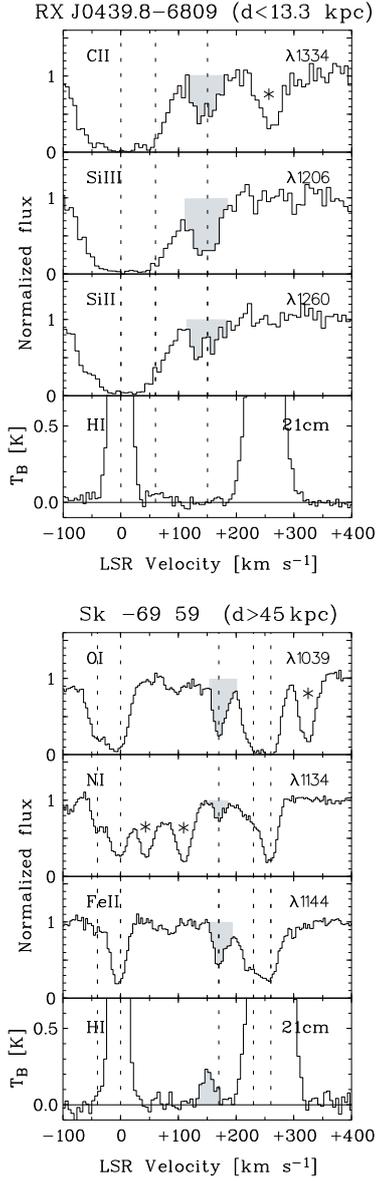}
 \caption{Absorption and emission profiles of different ions towards
 RX\,J0439.8$-$6809 and Sk\,$-$69\,59. HVC absorption/emission features
 are indicated with the gray-shaded area. The dotted lines indicate the
 LSR velocities of main absorption components in the Milky Way disk/halo and
 in the LMC. The star symbol indicates absorption from blending interstellar
 lines.
 }
 \label{figSpec}
 \end{figure}


 \begin{figure}[t!]
 \centering
 \includegraphics[width=6.0cm]{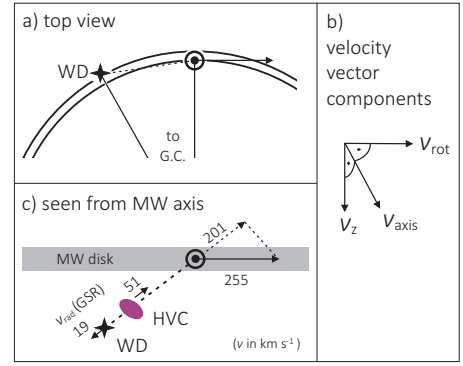}
 \caption{
 The geometry of Sun and RX\,J0439.8$-$6809 in the Galaxy shown schematically.
 {\it Panel a)}: view from above the Milky Way plane.
 {\it Panel b)}: decomposition of a space velocity in the Milky Way into the
 components $v_{\rm rot}$, the velocity parallel to the disk in the direction of
 circular rotation, $v_{\rm axis}$, the velocity component pointing
 (parallel to the disk) at the rotation axis of the Galaxy,
 and $v_{\rm z}$, the velocity perpendicular to the disk,
 with the outward direction having positive sign.
 {\it Panel c)}: the geometry as seen from the Milky Way rotation axis.
 Velocities are given in GSR (see Sect.\,4.1).
 The radial velocity of the star is +220\,\kms\ (Werner \& Rauch 2015).
 }
 \label{figVelVec}
 \end{figure}


\section{Properties of the high-velocity cloud}

\subsection{Location and kinematics}

With $l=279.9$, $b=-37.1$ and $d\leq 13.3$ kpc, the HVC's vertical
distance to the disk is limited to $z\leq 8.0$ kpc. The observed radial 
velocity, however, does not constrain the space velocity of the gas. 
To evaluate the space motion of HVCtwLMC 
we use the geometry of the sight line to RX\,J0439.8$-$6809 
as given in Fig.\,3.
First, we will consider velocities in the Galactic Standard of Rest (GSR), 
i.e., velocities with respect to the location of the Sun 
{\it without} the Galactic rotation. 
Assuming that the circular orbit speed of the Sun is 
$v_{\rm rot}=255$\,\kms\ (Reid et al.\,2014),
the component of the velocity of the Sun on the line of sight to the star is 
sin\,$(l)$\,cos\,$(b)\,v_{\rm rot} = 201$\,\kms\ 
(see Fig.\,3, panel\,c). 
The observed radial velocity of HVCtwLMC is +150\,\kms, 
so this gas has $+51$\,\kms\ in the GSR system in this direction
(Fig.\,\ref{figVelVec}, panel\,c).  

Another decomposition of the HVC space velocity vector is helpful.
We decompose it into three orthogonal components
(see Kaelble, de Boer \& Grewing 1985):
$v_{\rm rot}$, $v_{\rm axis}$, and $v_{\rm z}$ (see Fig.\,3, panel\,b).
The geometry of the Sun with WD is fortuitous in that,
even if HVCtwLMC would have a considerable $v_{\rm axis}$,
hardly any of that will show up in the observed $v_{\rm rad}$.
Therefore the $v_{\rm rad}{\rm (GSR)}$ of +51\,\kms\ can be understood
by just having $v_{\rm rot}= 64$\,\kms\ or by having $v_{\rm z}= -85$\,\kms.
In fact, the velocity components $v_{\rm z}$ and $v_{\rm rot}$ of HVCtwLMC
in the discussed geometry would have to combine
to result in $51$\,\kms $\simeq \cos(b) v_{\rm rot} - \sin(b) v_{\rm z}$
(note $v_{\rm z}$ has a negative sign for gas falling toward the disk).
Since essentially all neutral HVC gas in the halo is considered to be moving 
towards the MW disk, we conclude that $v_{\rm rot}$ most likely is less than 
64\,\kms, implying that HVCtwLMC is hardly participating in the Galactic rotation.
Because $v_{\rm axis}$ is, as said, undefined,
only these limits to the space motion of HVCtwLMC can be inferred.
See de\,Boer \& Savage (1983,1984) for more details
on this type of HVC velocity decomposition.

\subsection{Metallicity and ionization conditions}

The simultaneous presence of Si\,{\sc ii} and Si\,{\sc iii} towards
RX\,J0439.8$-$6809 provides strong evidence for the presence of 
ionized gas in HVCtwLMC. 
As discussed in Richter et al.\,(2015), it is likely that 
in diffuse circumgalactic environments these two ionization stages of Si 
trace predominantly complementary gas phases. 
The similar column densities of Si\,{\sc ii} and Si\,{\sc iii} 
(Table\,1) then indicate that there is at least as much ionized hydrogen 
in HVCtwLMC towards RX\,J0439.8$-$6809 as neutral hydrogen.
The column density of Si\,{\sc ii} suggests that the amount of neutral 
hydrogen is clearly below the upper limit as derived from H\,{\sc i} 21\,cm. 
Alternatively, if the gas would have solar composition, Si is depleted, 
but by at most 1\,dex.

For other abundance ratios we refer to data from the nearby sight line 
to Sk\,$-$69\,59 (separated by $\sim 2.5$ degrees), 
because for the sightline towards RX\,J0439.8$-$6809 
no H\,{\sc i} 21\,cm emission is detected in the LAB data 
and O\,{\sc i} $\lambda 1302.17$ absorption in the COS spectrum 
is contaminated by strong airglow lines. 
The most useful ion ratio to determine the overall metallicity of the gas 
is the O\,{\sc i}/H\,{\sc i} ratio. 
Neutral oxygen and neutral hydrogen have almost identical 
ionization potentials and in neutral gas regions 
both species are coupled through a strong charge-exchange reaction.
In addition, the $\alpha$ element oxygen is not significantly depleted 
into dust grains (e.g., Savage \& Sembach 1996).
We use H\,{\sc i} 21\,cm and O\,{\sc i} $\lambda 1039.23$ towards 
Sk\,$-$69\,59 (see Fig.\,2) to derive the metallicity of HVCtwLMC.
With the column densities given in Table\,1 and solar reference abundances 
from Asplund et al.\,(2009) we determine a metallicity/alpha abundance of 
[O/H]$=-0.63\pm0.05$ ($\sim 0.23$ solar), in good agreement with 
previous [O/H] estimates for the HVC in this direction (L09).

Another valuable indicator for the enrichment history of gas is the 
nitrogen abundance.
From the N\,{\sc i} column density we infer [N/H]$=-1.21\pm0.08$,
which is substantially lower than [O/H]. 
One reason for a low N/O ratio may be SN type II-dominated enrichment,
such as observed in other prominent HVCs (e.g., Richter et al.\,2001; 2013) 
and in many extragalactic absorption systems (e.g., Pettini et al.\,2008).
However, an alternative (and similarly plausible) explanation is that
the low N\,{\sc i} column density is a result of
ionization effects that become important for neutral nitrogen in
low-density environments (e.g., Fox et al.\,2013, their Fig.\,10).
The latter conclusion is supported by the shape of the 
N\,{\sc ii} $\lambda 1083.99$ line, 
which exhibits strong absorption at  $+170$\,\kms. Because of 
line saturation, we can only give a lower limit of log $N$(N\,{\sc ii})
$=14.25$ which does not help to further pinpoint the intrinsic
N/O ratio in HVCtwLMC without knowing the exact ionization conditions.
Similar arguments hold for iron, for which we determine an 
apparent abundance of [Fe/H]$=-0.73\pm0.06$, thus being close to
the derived [O/H] ratio. 
In predominantly neutral gas environments such a small difference 
between [O/H] and [Fe/H] would indicate very little dust depletion 
in the HVC in this direction. 
In HVCtwLMC, however, the observed Fe\,{\sc ii}/H\,{\sc i} ratio 
more likely reflects a considerable amount of H\,{\sc ii} that coexists 
with singly-ionized iron in mostly ionized gas regions (see also L09),
so that an intrinsically lower [Fe/H] due to dust depletion or 
nucleosynthetic effects cannot be excluded.

Summarizing, the analysis of [N/O] and [Fe/H] remains inconclusive 
with respect to the enrichment history and dust properties of the gas
because of the unknown ionization conditions.
The relatively low oxygen abundance of HVCtwLMC 
is in line with an extragalactic origin of the gas, similar
as for other HVCs such as Complex\,C and Complex\,A 
(Wakker et al.\,1999; Richter et al.\,2001).
The gas could also have originated in the Galactic halo,
being gas shed by metal-poor halo stars during their red-giant phase
(de Boer 2004), or from metal-poor gas in the outer Milky Way disk.
That the gas would be some form of outflow from the LMC is no longer 
tenable because of the above derived distance limit. 

\subsection{Total mass}


To provide an approximate mass limit of the HVCtwLMC we use the 
absorption statistics of O\,{\sc i} presented in L09 and, 
assuming that the distance of the cloud is $d=5$\,kpc 
and the angular size $A=100$\,deg$^2$, 
the angular size of the LMC. 
With these numbers, the physical size of the cloud is 
only 0.76\,kpc$^2$ and the neutral gas mass is small, 
$M_{\rm HVC}\approx 4\times 10^4 M_{\sun}$. 
Note that, depending on the 
ionization fraction of the gas and the spatial extent of the HVC beyond
the LMC boundaries, the total gas mass might be substantially higher than
this estimate. Additional absorption-line observations of 
QSOs that are located around the LMC would be very helpful to better constrain 
the size and mass of the HVCtwLMC.

\section{Conclusions}

Our observations demonstrate that the HVC in the direction 
of the LMC (as well as all other absorption components) 
are located in front of the star RX\,J0439.8$-$6809 at $<13.3$ kpc distance. 
They all are part of the Galaxy and do not belong to some LMC outflow.
HVCtwLMC thus is in line with other prominent HVCs for which reliable distance 
estimates exist and which are located within $20$ kpc distance, such as
Complex\,A, Complex\,C, the Cohen Stream, and the Smith Cloud
(van\,Woerden et al.\,1999; Wakker et al.\,2007, 2008; Thom et al.\,2006,2008).
The derived oxygen abundance of HVCtwLMC ($\sim 0.23$ solar) does not allow
us to draw firm conclusions about its origin; both scenarios, a Galactic and
an extragalactic origin, are in line with the observations.

The results presented here once again underline the importance 
of distance determinations of HVCs for our understanding 
of the spatial distribution of gas in the Milky Way halo.

\begin{acknowledgements}
T. Rauch was supported by the German Aerospace Center (DLR) under
grant 05\,OR\,1402. Some of the data presented in this paper were
obtained from the Mikulski Archive for Space Telescope (MAST).   
\end{acknowledgements}


\section*{References}

\begin{scriptsize}
Asplund, M., Grevesse, N., Jacques Sauval, A., \& Scott, P. 2009, ARA\&A, 47, 481\\
Danforth, C.W., Howk, J.C., Fullerton, A.W., Blair, W.P., \& Sembach, K.R. 2002, \\\indent ApJS, 139, 81\\
de\,Boer, K.S., \& Savage, B.D. 1983, A\&A, 265, 210\\
de\,Boer, K.S., \& Savage, B.D. 1984, A\&A, 136, L7\\
de\,Boer, K.S., Morras, R., \& Bajaja, E. 1990, A\&A, 233, 523\\
de\,Boer, K.S. 2004, A\&A, 419, 527\\
Fontana, A., \& Ballester, P. 1995, ESO Messenger, 80, 37\\
Fox, A.J., Wakker, B.P., Smoker, J.V., et al. 2010, ApJ, 718, 1046\\
Fox, A.J., Richter, P., Wakker, B.P., et al. 2013, ApJ, 772, 110\\
Fox, A.J., Wakker, B.P., Barger, A., et al. 2014, ApJ, 787, 147\\
Greiner, J., Hasinger, G., \& Thomas, H.-C., A\&A, 281, L61\\
Kaelble, A.,de\,Boer, K.S., \& Grewing, M. 1985, A\&A, 143, 408\\
Kalberla, P. M. W., Burton, W. B., Hartmann, D., et al. 2005, A\&A, 440, 775\\
Lehner, N., Staveley-Smith, L., \& Howk, J. C. 2009, ApJ, 702, 940\\
Morton, D.C. 2003, ApJS, 149, 205\\
Pettini, M., Zych., B.J., Steidel, C.C., \& Chaffee, F.H. 2008, MNRAS, 385, 2011\\
Putman, M.E., Peek, J.E.G., \& Joung, M.R. 2012, ARA\&A, 50, 491\\
Reid, M.J., Menten, K.M., Brunthaler, A., et al.\,2014, ApJ, 783, 130\\
Richter, P., de\,Boer, K.S., Bomans, D.J., et al.\,1999, Nature, 402, 386\\
Richter, P., Sembach,K.R., Wakker, B.P., et al.\,2001, ApJ, 559, 318\\
Richter, P. 2006, Reviews in Modern Astronomy 19, 31\\
Richter, P., Charlton, J.C., Fangano, A.P.M., Ben Bekhti, N.,
\& Masiero, J.R. 2009, ApJ,\\\indent 695, 1631\\
Richter, P., Fox, A. J., Wakker, B. P., et al. 2013, ApJ, 772, 111\\
Richter, P., Wakker, B.P., Fechner, C., et al.\,2015, A\&A, submitted\\
Savage, B.D. \& de\,Boer, K.S. 1981, ApJ, 243, 460\\
Savage, B.D.,\& Sembach, K.R. 1996, ARA\&A, 34, 279\\
Savage, B.D.,\& Sembach, K.R. 1991, ApJ, 379, 245\\
Staveley-Smith, L., Kim, S., Calabretta, M.R., Haynes, R.F., \&
Kesteven, M.J. 2003,\\\indent MNRAS, 339, 87\\
Smoker, J.V., Fox, A., \& Keenan, F.P. 2015, MNRAS, 451, 4346\\
Thom, C., Putman, M. E., Gibson, B. K., et al. 2006, ApJ, 638, L97\\
Thom, C., Peek, J. E. G., Putman, M. E., et al. 2008, ApJ, 684, 364\\
Wakker, B.P. \& van\,Woerden, H. 1998, ARA\&A, 35, 217\\
Wakker, B. P., York, D. G., Howk, J. C., et al. 2007, ApJ, 670, L113\\
Wakker, B. P., York, D. G., Wilhelm, R., et al. 2008, ApJ, 672, 298\\
Wannier, P., \& Wrixon, G. T. 1972, ApJL, 173, L119\\
Welty, D.E., Frisch, P.C., Sonneborn, G., \& York, D.G 1999, ApJ, 512, 636\\
Werner, K., \& Rauch, T. 2015, A\&A, in press\\
van\,Woerden, H., Schwarz, U.J., Peletier, R.F., Wakker, B.P., \&
Kalberla, P.M.W. 1999, \\\indent Nature, 400, 138\\

\end{scriptsize}

\end{document}